1

# Stability and normal zone propagation speed in YBCO coated conductors with increased interfacial resistance.

George A. Levin, Paul N. Barnes, Jose P. Rodriguez, Jake A. Connors, and John S. Bulmer

*Abstract*— We will discuss how stability and speed of normal zone propagation in YBCO-coated conductors is affected by interfacial resistance between the superconducting film and the stabilizer. Our numerical simulation has shown that the increased interfacial resistance substantially increases speed of normal zone propagation and decreases the stability margins. Optimization of the value of the resistance may lead to a better compromise between stability and quench protection requirements than what is found in currently manufactured coated conductors.

*Index Terms*—Coated Conductors, Normal Zone, Stability.

## I. INTRODUCTION

Superconducting wires in large scale applications have to satisfy two important criteria - they have to be stable during operation, but also allow integration of an effective quench protection scheme in the superconducting devices [1]. These two requirements compete with each other. Conventional low temperature superconductors operate at liquid helium temperatures, where their small heat capacity makes the current-carrying wires very unstable with respect to even a small amount of heat released in them. For that very reason, however, the speed of normal zone propagation (NZP) is high (several meters per second) which is beneficial for the rapid detection of the normal zone (NZ) and mitigation of the consequences of the quench. The situation is reversed in $YBa_2Cu_3O_{7-x}$ (YBCO) coated conductors [2]. Their main advantage over conventional superconductors is a high operating temperature (60-77 K) and the resultant high stability margins. The flip side is a low NZP speed which complicates quench protection [3-5].

We present the results of a numerical analysis of a model of NZP specialized to the architecture of the state of the art coated conductors. The main conclusion of this analysis is that the NZP speed can be increased up to an order of magnitude by increasing the interfacial resistance between copper stabilizer and the superconducting film. The concomitant reduction of the stability margin is substantial, but the wire retains an ability to recover on its own from a finite amount of locally injected heat.

After some of our preliminary results were presented elsewhere [6,7], we were advised by colleagues that a large body of work devoted to the effect of interfacial resistance on NZP in conventional superconductors had been done much earlier, with some of the publications predating the discovery of high temperature superconductivity [8-11]. This reminds us again of an ancient adage that "each new thing is a well-forgotten old one".

## II. MODEL

YBCO coated conductors are thin metal tapes in which an about 1 μm thick YBCO film is sandwiched between a metal substrate and the copper stabilizer [2,12]. The total thickness of the tape is about evenly divided between stabilizer and substrate and close to 100 μm. The standard width of the currently manufactured tape is about 4 mm. There is a very thin layer of material between the YBCO and copper that accounts for the resistance to the current exchange [13]. Common wisdom is that the resistance of this interface has to be as low as possible to ensure effective current transfer. If necessary, however, it can be increased by a chemical treatment or by changing the architecture of the wire. Here we treat the interface as an infinitesimally thin boundary between YBCO and copper with a finite resistance $\bar{R}$ [Ω cm$^2$].

Hereafter, we consider only the *direct transport current* condition. Three-dimensional (3D) equations of heat conduction in a thin tape-like composite wire, Fig. 1, can be reduced to the 2D (planar) or 1D linear model if the variation of temperature along the thickness of the wire is small in comparison with its variation along the plane of the wire. One can show that this criterion is satisfied if $d_i j_T \ll K_i(T_c - T_0)$, where $j_T$ is the thermal flux lifted from the surface of conductor, and $d_i$ and $K_i$, respectively, are the thickness and thermal conductivity of either the stabilizer or substrate material; $T_c$ and $T_0$ are the critical temperature and the operating temperature respectively ($T_c$ -$T_0$ ~20 K). For copper stabilizer $K_1 \approx$ 4-5 W/cm K and $d_1 \approx$ 40 μm. For a substrate metal like Hastelloy $K_2 \approx 7\times10^{-2}$ W/cm K and typically $d_2 \approx$ 50-100 μm [13]. As long as the heat flux from the surface is of the order of 1-10 W/cm$^2$ or less, the planar approximation is justified. The 150-300 nm thick buffer between the YBCO and substrate and even thinner YBCO-Cu interface also can be considered thermally transparent.

The 2D - {x,y} - heat conduction equation can be obtained by integrating the 3D equation along the thickness of the conductor (Fig. 1):

$$C\frac{\partial T}{\partial t} - K\Delta T = Q - 2K_0(T - T_0) \quad (1)$$

Here $C = C_1 d_1 + C_2 d_2$ is the combined specific heat, $K = K_1 d_1 + K_2 d_2$ is the effective thermal conductivity, and $Q = \int q(z) dz$ is the areal density of the internal heat sources. $K_0$ is the heat transfer coefficient across the insulation on the surface and $T_0$ is the ambient temperature which, in the absence of losses, is the operating temperature. We also neglect here for simplicity the temperature dependence of the material coefficients.

To obtain a closed system of equations for a simplified analysis we shall neglect the dissipation ($Q = 0$) in the sections of the wire that are subcritical, $J < J_c(T)$. This is an approximation because a certain amount of current flows in the stabilizer even in the subcritical sections [12]. The limits of this approximation are discussed below. We shall employ the Bean model approximation to describe the internal heat sources in the sections of the wire that are still superconducting, but where the transport current density exceeds the local critical current density, $J > J_c$(T) [9]. Hereafter the symbols $J$ denote the sheet current density [A/cm width]. There are three separate contributions originating from the stabilizer, the interface and the superconductor, respectively:

$$q = \frac{1}{\rho_1}\vec{E}_1^2 + \frac{(V_1 - V_s)^2}{\overline{R}}\delta(z) + J_s|E_s|\delta(z). \quad (2)$$

Here $\vec{E}_1$ is the electric field in the stabilizer and $\rho_1$ is its resistivity. $V_1$ is the local potential of the stabilizer and $V_s$ is the potential of the superconductor. The δ-functions account for the fact that two of the heat sources are concentrated in the volume much thinner than either the stabilizer or substrate. The heat dissipation in the substrate is negligible due to its large resistance [12] and the substrate contributes only to the thermal mass of the conductor (especially in light of the thin buffer layers).

Within the Bean model approximation the current through the superconducting film cannot exceed the critical current, so that $\vec{J}_s = \vec{n} J_c(T)$, where $\vec{n}$ is a unit vector in the direction of current. The excess of transport current must flow into the stabilizer. The condition of charge conservation $\nabla \cdot \vec{J}_s + j_z = 0$, where $j_z = (V_1 - V_s)/\overline{R}$ is the density of current across the interface, takes form

$$\vec{n} \cdot \nabla J_c = (V_1 - V_s)/\overline{R} \quad (3)$$

In the one-dimensional case that will be considered hereafter, Eq. (3) and its spatial derivative are given by

$$J_c^{/} = (V_1 - V_s)/\overline{R} ; \qquad \overline{R} J_c^{//} = E_s - E_1. \quad (4)$$

Here $J_c^{/} \equiv \partial J_c / \partial x$ and $J_c^{//} \equiv \partial^2 J_c / \partial x^2$. The direction of the current flow is taken to be positive, Fig. 1. Therefore, $E_s$ must be either positive or zero.

The power sources in the right-hand side of Eq. (1) have a different form in the three temperature regions: $T < T_1$, $T_1 \leq T \leq T_c$, and $T > T_c$, respectively. The current sharing temperature $T_1$ is defined by the condition $J_c(T_1) = J$. For $T < T_1$ (sub-critical regime) we take $Q = 0$. We want to emphasize that this is an approximation we have mentioned earlier. In fact when resistance of the interface is large, the current flows in the stabilizer even in the sufficiently "cold" sections where $J_c(T) > J$. The critical temperature $T_c$ is defined by the condition $J_c(T_c) = 0$. At temperatures above $T_c$ the transport current flows only in the stabilizer and $Q = \rho_1 J^2 / d_1$. All three heat sources in Eq. (2) contribute in the intermediate range of temperature, $T_1 \leq T \leq T_c$:

$$Q = \frac{\rho_1(J - J_c)^2}{d_1} + \overline{R}(J_c^{/})^2 + J_c\left(\frac{\rho_1(J - J_c)}{d_1} + \overline{R} J_c^{//}\right) H(E_s) \quad (5)$$

A unit step function $H(E_s)$ ensures that the electric field in the superconductor is either positive or zero. The necessity to introduce the step function "by hand" stems from the approximation mentioned above, when we neglected the precursor electric field propagating ahead of the normal zone [12] in the sections of wire where $T < T_1$. The step function ensures that the numerical solution does not "wander off" into unphysical territory where the heat source may become negative.

Rewriting Eq. (1) in dimensionless form will allow us to combine numerous material and experimental parameters into a smaller number of dimensionless quantities that determine different regimes of the NZP. A dimensionless temperature θ is defined as

$$\theta = \frac{T - T_1}{T_c - T_1}; \qquad J_c(\theta) = J(1 - \theta). \quad (6)$$

Here we also assumed for simplicity a linear dependence of the critical current on temperature, well justified for coated conductors [13]. The distances shall be measured in units of the thermal diffusion length $l_T = (D_T / \gamma)^{1/2}$, where $D_T = K / C$, and time in units of $\gamma^{-1}$, where the increment $\gamma = \rho_1 J^2 / d_1 C \Delta T$ determines the characteristic time required to warm an element of the conductor by $\Delta T = T_c - T_1$.

Equation (1) takes the following piece-wise form:

$$\frac{\partial \theta}{\partial \tau} - \frac{\partial^2 \theta}{\partial \xi^2} = 1 - \kappa(\theta - \theta_0); \quad \theta > 1. \quad (7)$$

$$\frac{\partial \theta}{\partial \tau} - \frac{\partial^2 \theta}{\partial \xi^2} = \theta^2 + r(\theta^{/})^2 + (1-\theta)(\theta - r\theta^{//})H(\theta - r\theta^{//}) - \kappa(\theta - \theta_0); \quad (8)$$
$$0 < \theta < 1.$$

$$\frac{\partial \theta}{\partial \tau} - \frac{\partial^2 \theta}{\partial \xi^2} = -\kappa(\theta - \theta_0); \quad \theta < 0. \quad (9)$$

Here $\tau \equiv \gamma t$ and $\xi \equiv x / l_T$. The cooling conditions are determined by the constants

$$\kappa \equiv 2 K_0 \Delta T d_1 / \rho_1 J^2 \text{ and } \theta_0 \equiv (T_0 - T_1)/(T_c - T_1) < 0. \quad (10)$$

The interface resistance enters into Eq. (8) as a combination

$$r \equiv \frac{\lambda^2}{l_T^2} = \frac{\overline{R}}{R_0}; \quad R_0 \equiv \frac{\rho_1 l_T^2}{d_1} = \frac{K(T_c - T_1)}{J^2}; \quad \lambda \equiv \left(\frac{\overline{R} d_1}{\rho_1}\right)^{1/2}. \quad (11)$$

Here λ is the current transfer length [12] that determines the distance over which the current exchange between the superconductor and stabilizer takes place. Although more complex, Eqs. (7-9) share common features with the standard Kardar-Parisi-Zhang (KPZ) equation [14]. In both cases the diffusion competes with the nonlinear growth term $\propto (\theta')^2$. The KPZ term corresponds to the heat released in the interface and its strength is proportional to the interface resistance.

The numerical solutions of Eqs. (7-9), subject to periodic boundary conditions, were obtained for an initial condition in the form of Gaussian temperature profile

$$\theta(\xi,0) = \theta_0 + (|\theta_0|+a)\exp\{-\xi^2/2\delta^2\}. \qquad (12)$$

All results presented here correspond to the same width of the initial profile $\delta=\sqrt{2}$. The value of the dimensionless ambient temperature is directly related to the transport current by virtue of the relationship (6):

$$\theta_0 = 1 - \frac{J_c^{(0)}}{J}, \qquad (13)$$

where $J_c^{(0)} \equiv J_c(T_0) > J$ is the critical current density at the operating temperature $T_0$.

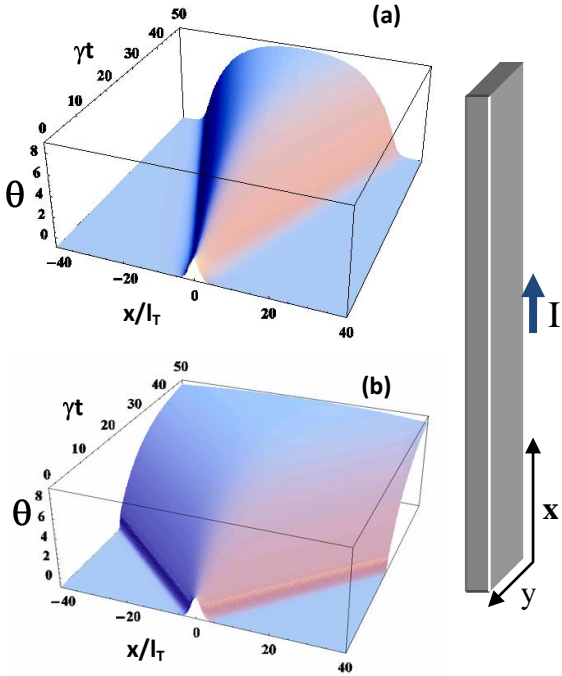

Fig. 1. Sketch of the conductor and two examples of the numerical solutions of Eqs. (7-9). Shown are the evolving temperature profiles $\theta(x,t)$. In both cases the initial condition (in the foreground) is the same, Eq. (12), with $\theta_0 = -1$, $a = 1.1$, and $\delta = 1.4$. The cooling constant $\kappa = 0.1$ is also the same. In (a) the interface resistance is negligible; In (b) $r = \bar{R}/R_0 = 1.5$.

### III. RESULTS AND DISCUSSION

#### A. Speed of Normal Zone Propagation

Figure 1 shows two examples, in the form of 2D plots, of the solutions $\theta(x, t)$ of Eqs. (7-9), subject to periodic boundary conditions and the initial condition given by Eq. (12). The normal zone evolves from the initial Gaussian perturbation with a width of the order of the diffusion length ($\delta =1.4$) and the peak temperature $a = 1.1$. This corresponds to the peak of the real temperature

$$T_{\max} = T_c + (a-1)(T_c - T_1) \qquad (14)$$

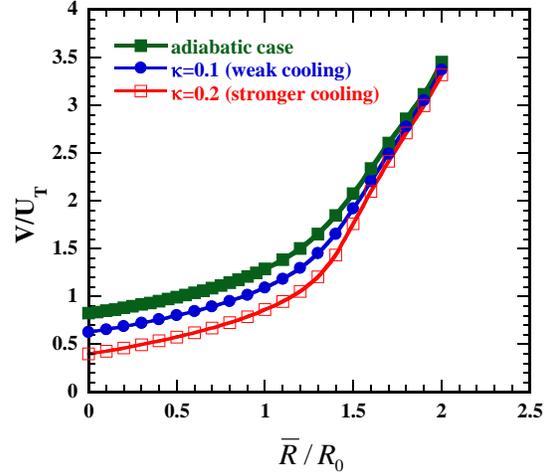

Fig. 2. The normal zone propagation speed as a function of interfacial resistance obtained from the solutions of Eqs. (7-9) similar to those shown in Fig. 1. The parameters of the initial condition are the same as in Fig. 1. The adiabatic case corresponds to $\kappa=0$. The value of $\theta_0= -1$ is the same for all cases.

The dimensionless ambient temperature $\theta_0 = -1$, that corresponds to the transport current density equal to 50% of the maximum critical current: $J = 0.5 J_c^{(0)}$. Under this condition the current sharing temperature $T_1$, defined by the condition $J_c(T_1) = J$, lies half-way between the operating temperature $T_0$ and the critical temperature, $T_1 = T_c + 0.5(T_c - T_0)$. The maximum temperature of the normal zone is determined by the cooling constant,

$$\theta_{NZ} = \theta_0 + \kappa^{-1}. \qquad (15)$$

The solutions shown in Figs. 1(a,b) correspond to $\kappa=0.1$, so that:

$$T_{NZ} = T_1 + 9(T_c - T_1). \qquad (16)$$

The only difference between the solutions in Figs. 1(a,b) is the value of the interfacial resistance. In Fig. 1(a) the resistance $r=0$. In Fig. 1(b) the interfacial resistance $\bar{R} = 1.5 R_0$ where $R_0$ is given by Eq. (11). The propagation speed determined by the slope $dx/dt$ at the level of $\theta=1$ $(T=T_c)$ increases drastically with increasing interfacial resistance.

In Fig. 2 the propagation speed determined from the solutions of Eqs. (7-9) is shown as a function of the interfacial resistance for three levels of cooling. The NZP speed is given by

$$V = U_T S(r, \theta_0, \kappa), \qquad (17)$$

where $U_T$ is a characteristic speed,

$$U_T \equiv l_T \gamma = (D_T \gamma)^{1/2} = \left(\frac{K \rho_1 J^2}{C^2 (T_c - T_1) d_1}\right)^{1/2}, \qquad (18)$$

and $S$ is a dimensionless speed enhancement factor that depends on interface resistance, the current redundancy $\theta_0$, Eq. (13), and cooling conditions. As Fig. 2 shows, the NZP speed, expressed in units of $U_T$, increases substantially as the resistance of the interface increases above the threshold $R_0$.

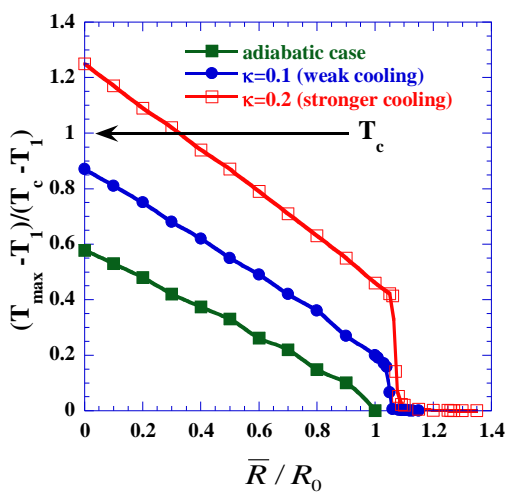

Fig. 3. Stability boundaries as functions of interfacial resistance for three levels of cooling. When the peak temperature of the initial temperature profile $T_{max}$ given by the parameter $a$ in Eqs. (12,14) exceeds the limit shown by the curves, the normal zone starts to propagate.

The material constants typical of coated conductors are as follows[13]: $K_1 \approx$ 4-5 W/cm K; $C_1 \approx$ 1.7 J/cm$^3$ K; $\rho_1 \approx 0.2 \times 10^{-6}$ $\Omega$ cm, and $d_1$ = 40 µm (copper stabilizer). $K_2 \approx 7 \times 10^{-2}$ W/cm K, $C_2 \approx$ 1.4 J/cm$^3$ K, and $d_2 \approx$ 50 µm (Hastelloy substrate). Let us also take the critical temperature, defined by the condition $J_c(T_c)$ =0, as $T_c \approx$ 90 K and the operating temperature $T_0$=65 K. As an example, we take the critical current density $J_c(T_0)$ = 200 A/cm, and the transport current density $J$ =100 A/cm (the corresponding value of $\theta_0$ = -1, as in Fig. 2). Then, the current sharing temperature $T_1 \approx$ 77 K. These parameters determine the increment $\gamma \approx$ 3.6 s$^{-1}$ and the thermal diffusivity $D_T \approx$ 1.4 cm$^2$/s, which gives the thermal diffusion length $l_T \approx$ 6 mm. Correspondingly, the characteristic speed $U_T = l_T \gamma \approx$ 2.2 cm/s. Considering that for low interface resistance the NZP speed is approximately ( 0.5-0.7)$U_T$, see Fig. 2, these estimates agree well with the results of the experiments[4,5] that found the NZP speed in coated conductors of the order of 1 cm/s.

The characteristic interface resistance above which we can expect to see a substantial increase in propagation speed is $R_0=\rho_1(l_T)^2/d_1 \approx 18 \times 10^{-6}$ $\Omega$ cm$^2$. The typical interfacial resistance of currently manufactured coated conductors is of the order of 50 n$\Omega$ cm$^2$[15]. Therefore, the interface resistance has to be increased by at least two orders of magnitude in order to observe its effect on NZP speed (see Fig. 2).

*B. Stability*

Stability is defined as an ability of a current-carrying conductor to dissipate a certain amount of heat and return to normal operation on its own. The NZP speed and stability margins usually anti-correlate in superconducting wires. Our analysis shows that the effect of interfacial resistance is no exception from this rule. With increasing resistance the NZP speed increases, but the ability of the coated conductor to recover from a perturbation diminishes. We determined the margins of stability by solving Eqs. (7-9) with the initial condition (12) and finding a peak temperature, determined by the constant $a$ in Eqs.(12, 14), at which the normal zone starts to propagate. The width of the perturbation was kept constant, the same as in the data in Figs. 1 and 2. If the peak temperature is below a respective curve in Fig. 3, the initial temperature profile dissipates and the normal zone does not propagate. For the peak temperatures above the curve, the normal zone expands.

As long as the interfacial resistance $\overline{R}$ is below the threshold $R_0$ the initial temperature perturbation can dissipate without triggering a NZP even if the peak temperature exceeds the current sharing temperature, so that a fraction of the transport current initially flows through the stabilizer. Even when the peak temperature exceeds the critical temperature the conductor remains stable if the cooling rate determined by the constant $\kappa$ is sufficiently large (the arrow in Fig. 3). With increasing resistance the ability of the conductor to recover strongly declines and for $\overline{R} > R_0$ a perturbation with a peak temperature even slightly in excess of $T_1$ gives rise to a NZP. It is important to emphasize that at $\overline{R} > R_0$ the conductor is still able to absorb a finite amount of heat and remain stable, but only as long as no current is diverted into the stabilizer (within our model it means $T_{max} < T_1$).

IV. CONCLUSIONS

By increasing the interfacial resistance by about two orders of magnitude over that in currently manufactured coated conductors we can expect to see a substantial increase in speed of the NZP. A side effect of this is a decrease in stability margins.

ACKNOWLEDGMENT

We thank R. Mints, A. Gurevich, and R. Duckworth for their interest in this work and useful references to previous publications on the subject.